# Searching for low-mass stars with magnetically-induced hyper-inflated radii

D. J. Mullan & J. MacDonald

Department of Physics and Astronomy, University of Delaware, Newark DE 19716

**Abstract**

Precise empirical estimates of stellar radii have revealed that the radii of certain low-mass stars are inflated relative to stellar structure predictions: the largest inflations occur in magnetically active stars. Theoretically, the radii of magnetically active stars are in some cases found to be "hyper-inflated" to roughly double the radius of a non-magnetic star with equal mass. Here we ask, do data exist which could allow us to search for empirical evidence in support of hyper-inflated stars? A photometric study of 44 eclipsing binaries in the Kepler field by Cruz et al. may help us in our search. The Cruz et al. study, although subject to large uncertainties, hints at the presence of hyper-inflation in some of the 88 stars in their sample. Their data enable us to set theoretical limits on the maximum strength $B_c$ of magnetic fields inside their sample stars. According to our magneto-convective model, the average empirical inflations found from analysis of the Cruz et al. data can be replicated if $B_c \approx 10$ kG inside stars with masses greater than ~ 0.6 M$_\odot$. On the other hand, in stars with masses less than ~ 0.4 M$_\odot$, our model predicts that the average empirical inflations of the stars may approach hyper-inflated status. Such stars may require significantly stronger internal fields, i.e. $B_c \approx 100$-300 kG. High-resolution spectroscopy of the Kepler binaries could help to confirm or refute our conclusions.

## 1. Introduction

Empirical data on the radii of low mass stars indicate that some such stars are "inflated" in radius above the predictions of standard stellar models (e.g. Leggett et al. 2000), especially if the stars exhibit features of magnetic activity. Mullan & MacDonald (2001: MM01) computed magnetic stellar models using a criterion derived by Gough & Tayler (1966: GT66) for the onset of convection in terms of the strength of the vertical magnetic field. Specifically, the GT66 criterion for magneto-convective onset is written as $\nabla > \nabla_{ad} + \delta$ where $\delta = B_v^2/(B_v^2 + 4\pi\gamma P)$ is a "magnetic inhibition parameter". In the formula for $\delta$, $B_v$ is the vertical component of the local magnetic field, $P$ is the local gas pressure and $\gamma$ is the local ratio of specific heats. The principal uncertainty in applying the GT66 criterion is to specify how $\delta$ varies with the radial coordinate inside the star. The simplest assumption, $\delta(r)$ = const., leads to fields of 10's of MG in low-mass stars: such fields are impossibly strong to be generated inside such stars. As an alternative, we settled on maintaining $\delta(r)$ = const. only in the *outer layers* of the star: this led to a field strength which increases inward from the surface. A "ceiling" is imposed (by *fiat*) on the field when, as one moves deeper into the star beneath the surface, $B_v$ first reaches the selected ceiling value $B_c$ at a depth of $z_{ceil}$. At deeper levels, i.e. at $z > z_{ceil}$, the field strength is held constant (again, by *fiat*) at the value $B_c$.

Using this "GT66 criterion" approach, we found that many empirical values of relative radius inflation (= $\Delta R/R$) as large as ~ 20-30% can be replicated by selecting $B_c$ = 10 kG (e.g. MacDonald & Mullan 2017: MM17).

Independent of the magneto-convective (GT66) criterion, the occurrence of cool star spots on the surface of a star can also lead to radius inflation (e.g. MacDonald & Mullan 2012). It is a matter of interest to compare these two distinct approaches (i.e. magneto-convective versus star spot) to determine if one might be empirically preferable to the other.

## 2. Predicting upper limits on radius inflation in the magneto-convection model

The main purpose of the paper by MacDonald & Mullan (2024a: MM24a) was to examine the following question: does there exist any theoretical *upper limit* on the amount of radius inflation in low-mass stars? In other words, is there any reliable evidence for "hyper-inflation", i.e. stars with radii inflated by significantly more than $\Delta R/R \approx$ 20-30%? (The latter values are the largest inflations so far reported in the literature (see MM24a, their Fig. 1.) In the context of the MM24a stellar models based on the GT66 criterion, the theoretical answer to the question of hyper-inflation is (in certain cases) *Yes*.

To be specific, MM24a constructed stellar models in which the strength of the ceiling field $B_c$ was assigned to have one of three values: 10 kG, or 100 kG, or 1 MG. The reason for selecting these values is as follows. The lower limiting choice on $B_c$ was based on the successful fits to empirical radius inflations in a sample of almost 20 low-mass stars reported by MM17: in view of our success in fitting so many main sequence stars with $B_c$ = 10 kG, we consider it unlikely that stars which are magnetic enough to exhibit such macroscopic features as spots and/or flares would have $B_c$ < 10 kG. The largest choice of $B_c$ was based on the upper limit of field strength which can be stably confined in a solar mass star (Browning et al. 2016). And the choice of 100 kG was made as an intermediate value between the upper and lower limits just mentioned.

MM24a found that, in stars where $B_c$ = 10 kG, the theoretical value of the fractional radius inflation $\Delta R/R$ exhibits a clear maximum as a function of mass: $(\Delta R/R)_{max} \approx$ 90% for stars of mass 0.7 $M_\odot$. In stars with masses > 0.7 $M_\odot$, as well as in stars with masses < 0.7 $M_\odot$, the GT66 models with $B_c$ = 10 kG predicted $\Delta R/R$ values which were ≤ 80%. In stars with $B_c$ = 100 kG, MM24a once again found a clear maximum in the value of $\Delta R/R$ as a function of mass: $(\Delta R/R)_{max} \approx$ 130% in stars of mass 0.5 $M_\odot$. Once again, in stars with masses > 0.5 $M_\odot$, as well as in stars with masses < 0.5 $M_\odot$, the GT66 models with $B_c$ = 100 kG predicted $\Delta R/R$ values which were ≤ 130%. Finally, in stars with $B_c$ = 1 MG, the models did *not* indicate evidence for any maximum value of $\Delta R/R$: instead, the value of $\Delta R/R$ was found to continue to increase as mass increases, reaching a value of $\Delta R/R \approx$ 350% in stars of mass 0.9 $M_\odot$ (the maximum mass for which a converged model could be computed).

In view of these results, we suggest a definition of "hyper-inflation" might be appropriate for stars with radii which exceed the standard radii by factors of 90-130%. That is, a "hyper-inflated" star has a radius which may be roughly double the radius of a non-magnetic star with the same mass.

From a physics perspective, it is worthwhile to identify the predominant feature which gives rise to maximum radius inflation in models with $B_c$ = 10 and 100 kG (see MM24a, Section 5.4). Briefly, in the context of GT66 magneto-convective models, the presence or absence of a maximum value of relative radius inflation $\Delta R/R$ depends on two critical depths inside the star: $z_{ceil}$ (defined in Section 1 above) and $z_{base}$ (the depth of the base of the convective envelope). On the one hand, if $z_{ceil}$ lies *inside* the convective envelope (i.e. if $z_{ceil} < z_{base}$) for all values of the magnetic inhibition parameter $\delta$ (see Section 1), then $\Delta R/R$ *will* exhibit a maximum at a certain value of $\delta$. On the other hand, if $z_{ceil}$ lies *deeper* than the base of the convective envelope (i.e. if $z_{ceil} > z_{base}$), then $\Delta R/R$ will *not* exhibit a maximum value, but will increase monotonically as $\delta$ increases.

### 3. Empirical evidence for radius inflation

Are there empirical data which allow the upper limits on $\Delta R/R$ predicted by MM24a to be tested?

Observational evidence indicating how the amount of radius inflation $\Delta R$ varies as a function of the stellar mass has been available for some time, e.g. Lopez-Morales (2007), and Sebastian et al. (2023). In figures which are presented in those papers, visual inspection provides two significant pieces of information: (i) the smallest values of $\Delta R$ are observed in stars with the smallest masses, specifically, masses ≤ 0.3 - 0.35 $M_\odot$; (ii) empirical $\Delta R$ values which are definitely larger than those in the lowest mass stars are found to occur in stars with "intermediate" masses between 0.4 and 0.6 $M_\odot$. In the case of stars in class (ii), fractional radius inflations are reported to have numerical values of $\Delta R/R$ which can be as large as ~ 20-30%.

So far, the empirical items (i) and (ii) in the preceding paragraph are satisfactorily consistent with the results of MM24a discussed in Section 2 above.

Now we proceed further and raise a broader problem: is there any reliable empirical evidence for "hyper-inflated" stars?

### 4. Comparison between the largest empirical values of radius inflation and the theoretical limits predicted by magneto-convection

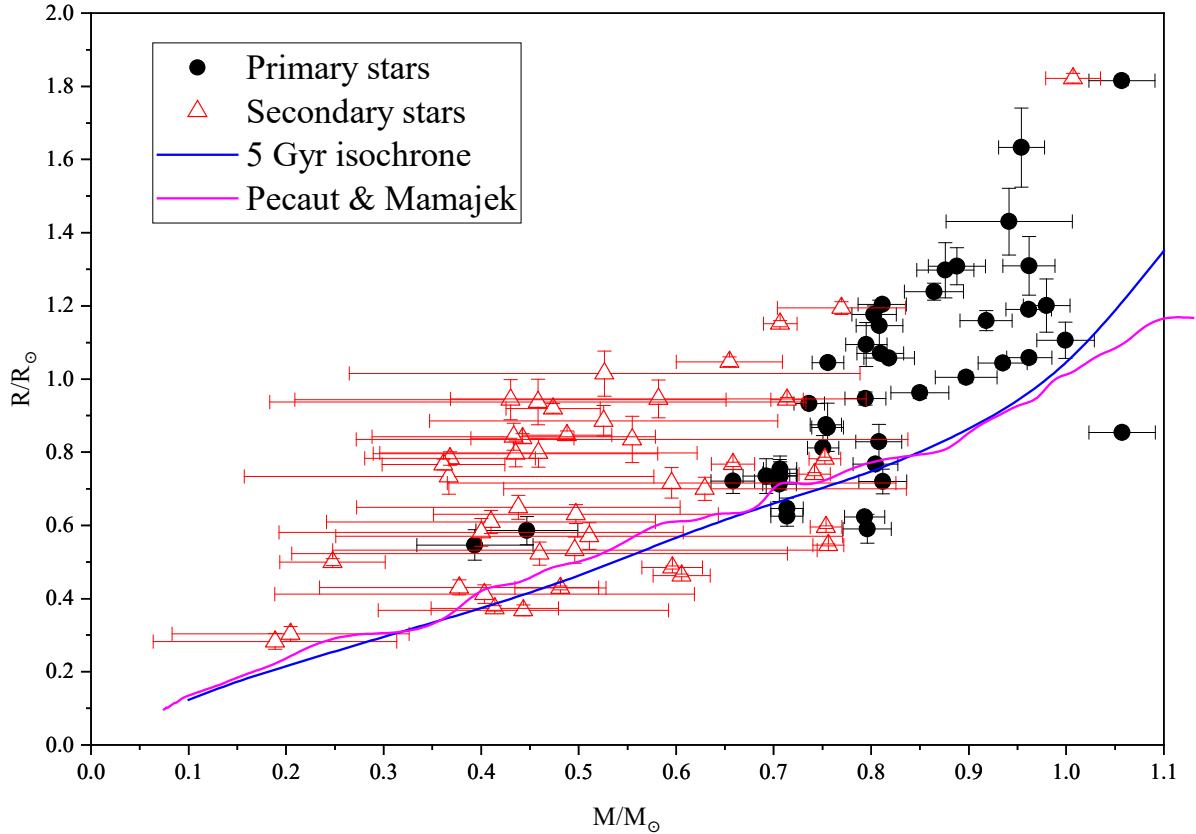

Figure 1. Stellar radii plotted against stellar mass for the sub-sample B of Cruz et al. (2022). The magenta line shows the spline interpolated data from the table of Pecaut & Mamajek (2013) and the blue line shows the 5 Gyr isochrone from solar composition models of standard (i.e. *non-magnetic)* stars.

Cruz et al. (2022a, hereafter C22) have reported on an analysis which allowed them to extract stellar temperatures, radii, and masses by using photometric data for 88 components of 44 detached eclipsing binaries in the Kepler field. (In the notation of C22, the 88 components belong to their "sub-sample B".) Using evolutionary models computed by Bressan et al. (2012), C22 selected dwarf star models with $T_{eff}$ in the range 2303-5991 K and they created a grid of synthetic model binaries containing almost 200,000 entries, for each of which they computed 10 different colors based on the magnitudes reported by PanSTARRS (*g,r,z,i,y*) and by 2MASS (*J,H,K*). Using the reported magnitudes, Cruz et al. first constructed the following 7 standard colors with adjacent passbands for each of their targets: (*g – r*), (*r – i*), (*i – z*), (*z – y*), (*y – J*), (*J – H*) and (*H – $K_s$*). In addition, they added three redder colors adopted by Garrido et al. (2019) in order to better identify the binaries which consist of two main sequence stars: (*r – J*), (*r – H*) and (*z – $K_s$*). The empirical data in all 10 colors were then interpolated into the grid to determine for each star its $T_{eff}$. C22 then made use of a table of $T_{eff}$ and masses compiled by Pecaut & Mamajek (2013)[1] to extract a "photometric mass" for each star. With $T_{eff}$ values in hand, the light curve of each binary was analyzed photometrically to extract the sum of the radii of the two stars and the ratio of the two radii in units of binary separation: Combining these quantities with the binary separation determined from Kepler's Third Law, C22 obtain "photometric radii" of primary and secondary components.

Even a casual inspection of Fig. 1 indicates that the number of stars lying *above* the 5 Gyr (non-magnetic) isochrone is much larger (by a factor of order 10) than the number of stars lying *below* that isochrone. That is, most of the stars in the C22 sample exhibit *inflated radii* relative to models of non-magnetic stars. Moreover, among the less massive stars (≤ 0.4 M$_\odot$), a number of stars have radii which appear to exceed the 5 Gyr isochrone by a factor of about 2. And even among the more massive stars (≥ 0.6 M$_\odot$), a number have radii which exceed the 5 Gyr isochrone by a factor of 50% or more. These inflations are well above the inflations of (20-30)% which already exist in the literature (see Section 3 above).. In our opinion, excesses in radii by ≥50%, if verified, could qualify for the label of "hyper-inflation".

However, this "casual" conclusion must be treated with caution: a number of objects in C22's sub-sample B, particularly the secondary stars, have large uncertainties in $T_{eff}$, which lead to large uncertainties in mass. Cruz et al. (2022b) have pointed out that these large uncertainties can lead to misleading conclusions *in the case of certain individual systems*. In view of this warning, rather than singling out individual systems for discussion, we use here a statistical approach to determine how the degree of radius inflation correlates *on average* with stellar mass. For each binary system, we draw $T_{eff}$ values for each component from a normal distribution with the mean and standard deviation given by the C22 photometric analysis. We then interpolate in the Pecaut & Mamajek (2013) table to find the component masses. We discard masses lower than the hydrogen burning limit of 0.075 M$_\odot$ (Chabrier et al. 2023). The sum of the component masses is combined with the orbital period to find the orbital separation. The ratio of the sum of stellar radii to orbital separation and the ratio between stellar radii determined from the C22 eclipse analysis are then used to determine the individual "photometric" stellar radii $R_{ph}$. Our results for the stellar masses and radii (including uncertainties) are shown in Figure 1. It is noteworthy that, although the uncertainties *in mass* for many of the secondary stars are quite large, nevertheless, the error bars *in radius* are relatively small. There is a clear physical reason for this feature: the radii depend on the sum of the masses for each system to only the 1/3 power. Here the amount of (relative) radius inflation is defined to be the ratio of the radius determined by this technique $R_{ph}$ to that from interpolation in the radii $R_{tab}$ tabulated by Pecaut & Mamajek. For each binary, we repeat this process 10,000 times to determine the mean and standard deviation of the radius inflation for mass binned from 0.1 to 1.1 M$_\odot$ in intervals of 0.1 M$_\odot$. In terms of the *fractional* radius inflation mentioned in Section 1 above, $\Delta R/R$, the latter quantity is defined as $(R_{ph} - R_{tab})/R_{tab}$.

---

[1] Pecaut & Mamajek regularly update their tabulated values, which are available at
https://www.pas.rochester.edu/%7Eemamajek/EEM_dwarf_UBVIJHK_colors_Teff.txt

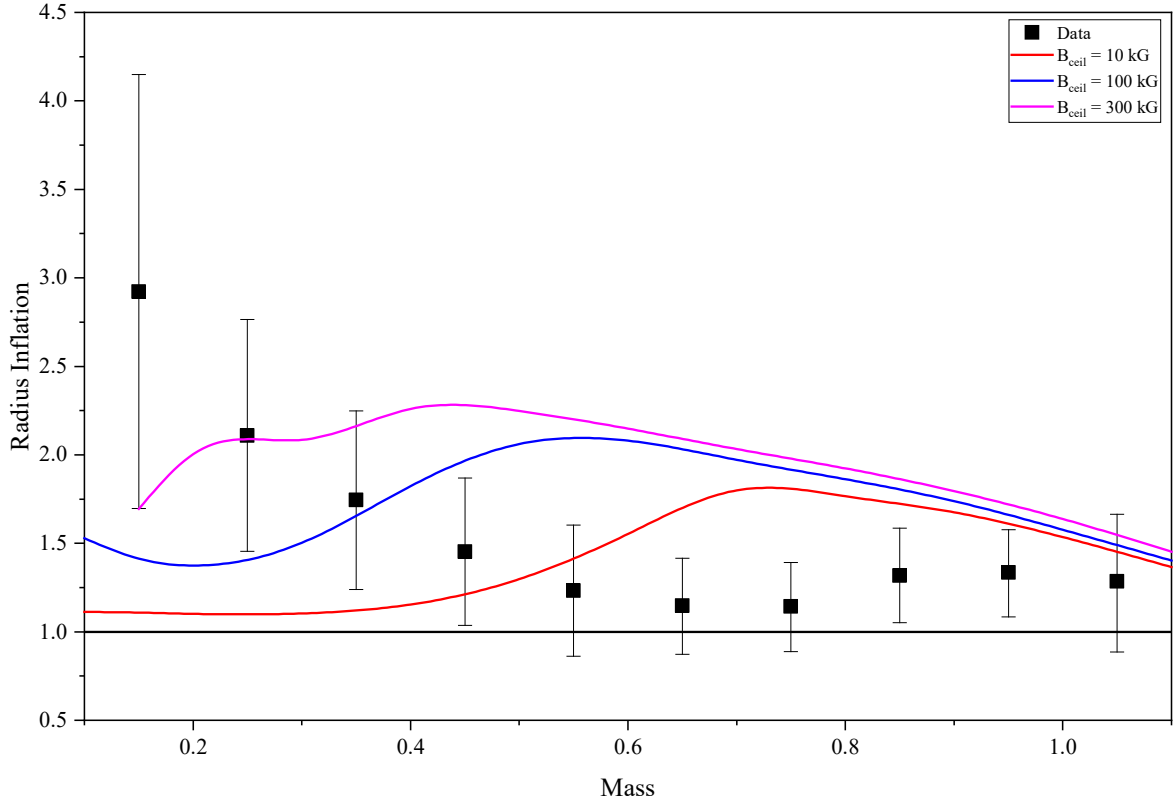

Figure 2. Radius inflation (defined as the ratio of inflated stellar radius to standard stellar radius for a star of equal mass) derived from statistical analysis of Cruz et al. (2022) data compared to predictions from our magneto-convection models. The horizontal line corresponds to no inflation.

We compare radius inflation predictions *from our magneto-convection models* with the empirical radius inflations determined from our statistical analysis of the C22 data in Figure 2. For the model predictions, the radius inflation plotted in Fig. 2 is defined to be the ratio of the (inflated) radius of the magnetic model to the radius of the non-magnetic model with the same mass.

Inspection of Fig. 2 shows that for stars with masses ≤ 0.4 $M_\odot$, the *average* empirical radii are inflated by values ranging from 50% to ≥ 100%. According to our proposed definition, such stars may qualify for the title "hyper-inflated".

The colored lines in Fig. 2 show the predicted radius inflation for 5 Gyr isochrones of our magneto-convective models for three values of $B_c$. We have chosen $\delta = 0.4$ as a representative value of the magnetic inhibition parameter (see MM24a for the effects of choosing a different value for $\delta$).

We see that, according to our magneto-convection models of stars with masses greater than about 0.6 $M_\odot$, the mean empirical inflations determined from the C22 data can be achieved with $B_c \leq 10$ kG. A value for the "ceiling" field $B_c$ of no more than 10 kG has already emerged in our earlier work (MM17 as replicating empirical radii in some 20 stars on the lower main sequence (see Section 2 above). This shows that the *average* radius inflations for Kepler stars with masses ≥ 0.6 $M_\odot$ are quite consistent with the inflations that MM17 accounted for in their independent sample of 20 stars.

However, when we examine stars with smaller masses, a different value of the ceiling field emerges: $B_c$ must be of order 100 kG to replicate the *average* inflations in stars of mass between ~0.4 and ~0.6 $M_\odot$. And for stars of mass less than ~0.3 $M_\odot$, an even stronger ceiling field, $B_c$ = 300 kG, is required to replicate the *average* inflations.

*Inclusion of a "photometric" caveat:* in their photometric analysis, C22 made use of evolutionary tables (Bressan et al. 2012) which were computed for "*standard" (non-magnetic) stellar models*. However, we are dealing here with *magnetic* stars, and for these, the surface temperatures can be altered from the Bressan et al. values by magnetic effects. (e.g. MM01) As a result, the 10-color system which is suitable for deriving photometric properties of non-magnetic stars needs to be corrected if the colors are to be applied to magnetic stars. Specifically, MacDonald and Mullan (2022) pointed out that the "photometric masses" derived by C22 are actually *lower limits* on the actual masses. But the corrections in mass are highly non-linear: stars with masses of 0.65-0.71 $M_\odot$ need the smallest corrections, and these are the stars which are predicted to undergo the maximum inflation (MacDonald & Mullan 2024b). On the other hand, stars with masses of 0.13-0.39 $M_\odot$ require larger corrections in "photometric mass". Thus, some of the estimates of radius inflation inferred from analysis of the C22 data might need to be reduced, especially in the range of masses which lie close to the transition to complete convection on the main sequence. We will address this issue in more detail in Section 6 below.

5. **Are there theoretical limits on radius inflation due to star-spots?**

As mentioned in Section 1 above, star spots and magneto-convection have both been found to be possible mechanisms to replicate radius inflation. It is natural to ask: are there also limits on the maximum amount of radius inflation predicted by star spot models?

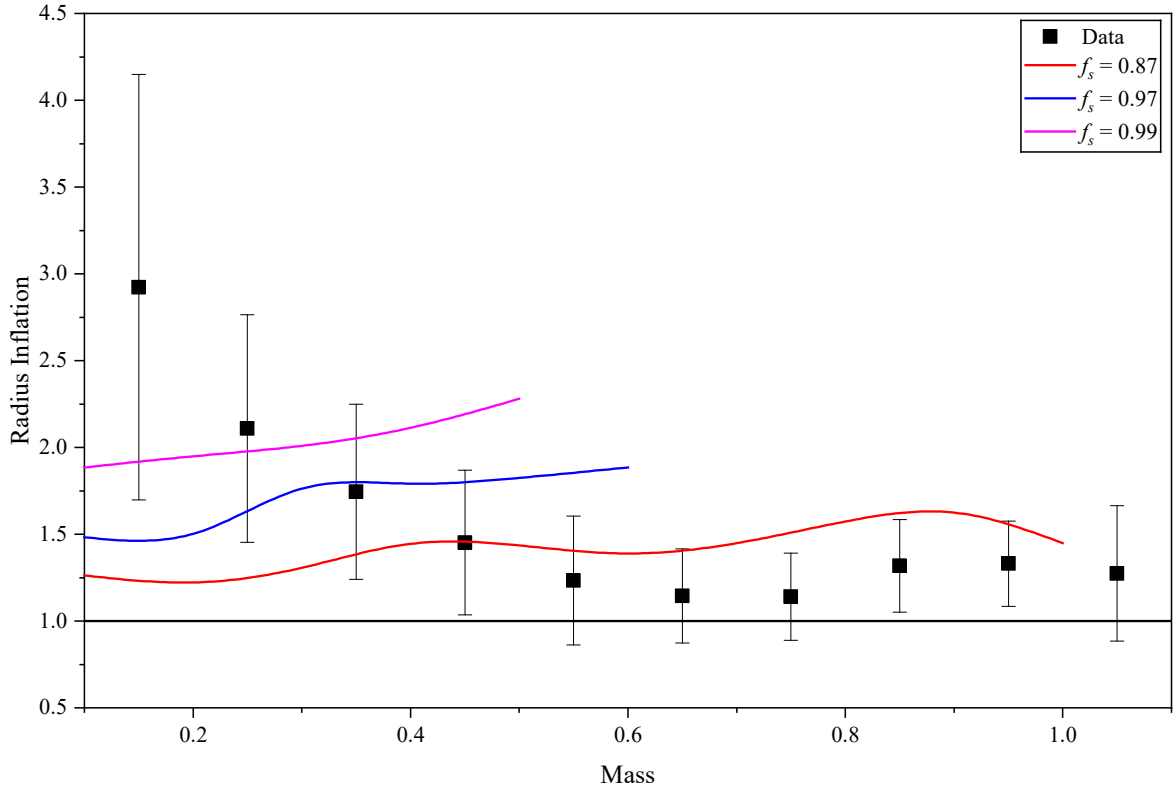

Figure 3. Radius inflation (i.e. ratio of inflated radius to standard radius at a given stellar mass) from statistical analysis of Cruz et al. (2022) data compared to prediction from our dark starspot models. The horizontal line corresponds to no inflation.

Using the model for the effects of completely dark spots (i.e. with spot temperatures of zero) described by MacDonald & Mullan (2012) and comparing with our statistical analysis of the C22 data, we get the results shown in Figure 3. We find the average radius inflations for stars with masses greater than about 0.5 $M_\odot$ are consistent with dark spot coverage factors of 87% or less. For lower mass stars, larger dark spot coverage factors are needed to explain the mean observed inflations: a coverage of 99% is needed for masses less than ~0.3 $M_\odot$.

We stress that although the general trends of radius disagreement at masses >0.4 $M_\odot$ in Figure 3 seem to be well traced by starspot models, these trends should be viewed with caution because these estimates of the spot coverage factors are based on the assumption that the starspots are *completely dark*. If we relax this condition, we might allow the spots to have finite temperatures of 80% of the unspotted surface temperature (such as was assumed by Torres et al. [2022] in their study of the EPIC 219511354 secondary). But in such a case, we find that fits to the empirical inflations from our statistical analysis of the C22 data are impossible to achieve.

Although both spots and magneto-convective models can be made to fit data associated with stellar inflation, we consider magneto-convective models as being more physically realistic. The reason is that magneto-convective models calculate the structural changes which occur in all regions throughout the star, whereas helioseismology analysis indicates that spots typically involve only those layers of the star which lie close to the surface, within 1-2% of the solar radius (e.g. Gizon et al. 2009).

## 6. Effects of systematic errors in temperatures of the secondaries

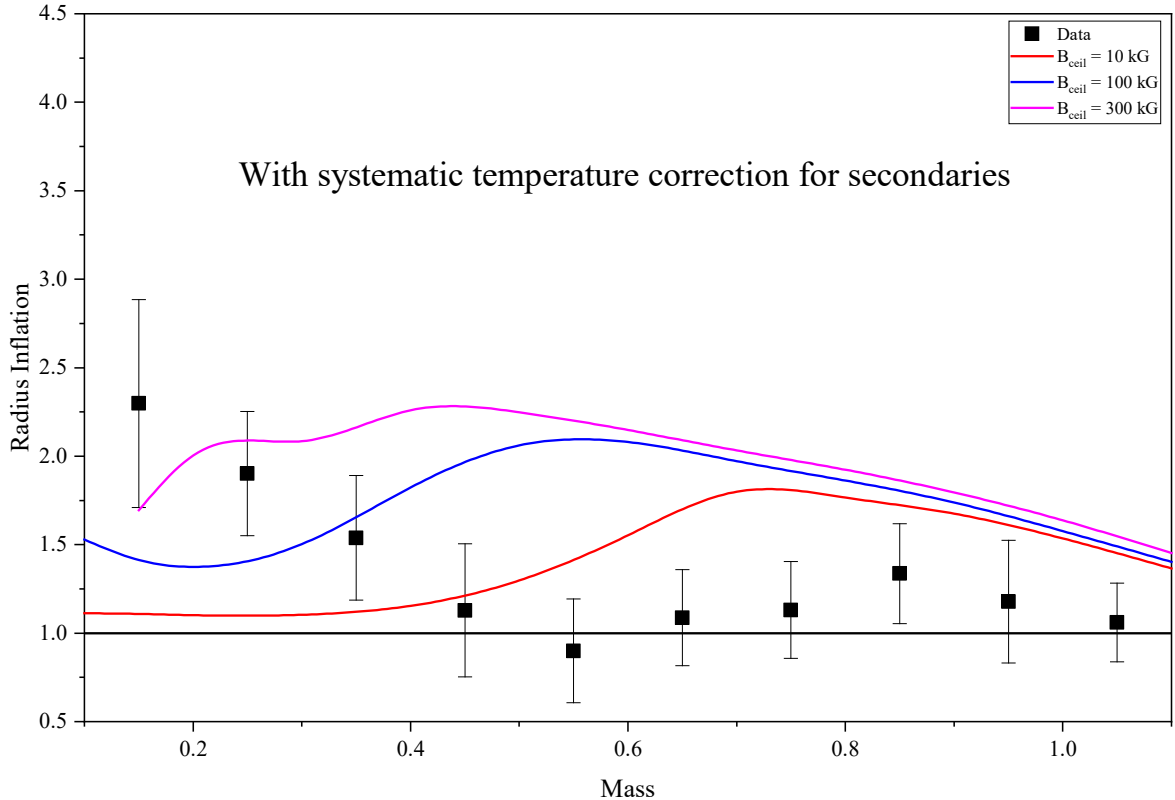

Figure 4. As in Figure 2 except that the effective temperatures of the secondaries are modified to explore the possible impacts of systematic differences between photometric and spectroscopic temperatures.

Inspection of fig. 2 in C22 indicates that there might be systematic differences in the secondaries' $T_{eff}$ values obtained by C22 from their analysis of *photometric* colors and values in the literature which have been mainly obtained by *spectroscopic* analysis. The spectroscopic temperatures are typically ~20% *larger* than the photometric values, which imply that the secondaries are more massive than the photometric masses and would have lower inflation. To quantify the possible effects of systematic temperature differences, we have performed a linear regression fit between the spectroscopic and photometric $T_{eff}$ values for the secondaries in the 'Correctly assigned V+V systems' in Table 1 of C22. We have used the linear regression fit to 'correct' the photometric temperatures in our inflation analysis. In addition, we have reduced the uncertainty in the temperatures by a factor of 3 to reflect the greater accuracy of spectroscopic temperature determinations. The results of this experiment are shown in Figure 4. As expected, the derived inflations are now found to be smaller, particularly for masses less than 0.6 M$_\odot$. But our main conclusion still emerges as before, namely, the secondaries with masses less than ~0.3 M$_\odot$ are on average 'hyper-inflated', i.e. they have radii a factor of ~2 greater than a non-magnetic star of the same mass.

## 7. Conclusions and discussion

Increasing numbers of low-mass stars have been identified as having radii which are larger than standard stellar structure calculations predict. The occurrence of "radius inflation" is especially prevalent among stars which exhibit symptoms of magnetic activity. In view of this, we have been applying a model of magneto-convection in order to replicate the empirical inflations.

In 2024, we used our model to compute theoretically the *maximum* amounts of magnetically-induced inflation in stars of various masses in the presence of certain upper limits on the magnetic field strength inside the star (see MM24a). The upper limits we selected for the internal field strength $B_c$ were 10 kG, 100 kG, and 1 MG: the latter value is the strongest field which can exist stably inside a low-mass star (Browning et al. 2016). In the case of 10 kG fields, we found that the maximum value of the fractional radius inflation would be $\Delta R/R \approx 90\%$ in a star of mass 0.7 M$_\odot$. (We suggest in the present paper that the label "hyper-inflation" would be suitable for such large inflations.) In the presence of 100 kG fields, $\Delta R/R$ was predicted to reach about 130%. And in the limit of 1 MG, relative inflations of as much as 350% were predicted.

The goal of the present paper is to search for empirical evidence (if such exists) in support of these predictions of "hyper-inflation".

In 2022, C22 published empirical results that permit estimates of $\Delta R/R$ for some 88 stars in Kepler data: in some of their stars, they concluded that $\Delta R/R$ could be as large as $\approx 100\%$. (**See Section 4 above for a *caveat* regarding this conclusion.**) In the present paper, we have performed a statistical analysis of the C22 data to determine the mean and standard deviation of the radius inflation for stellar masses between 0.1 and 1.1 M$_\odot$ binned in intervals of 0.1 M$_\odot$. We find that for stars of mass $\geq 0.6$ M$_\odot$, maximum internal fields of 10 kG can replicate the mean inflations found from our statistical analysis of the C22 data. This is consistent with earlier work of ours which suggested that a maximum field strength of 10 kG inside a star can replicate the inflations which have been reported in some 20 stars on the lower main sequence (MM17). However, we arrive at a different conclusion as regards stars of mass less than ~ 0.3 M$_\odot$: for these stars, we find that the stellar radii are large enough to qualify for the label of "hyper-inflation". Moreover, we find that the maximum field strength inside such low-mass stars must be as large as 300 kG.

If these results can be confirmed, they will suggest that, other things being equal, a dynamo in a completely convective star may generate significantly stronger fields than a dynamo in a star consisting of a radiative core plus a convective envelope.

**Acknowledgements**

J.M. acknowledges partial support from the Delaware Space Grant Consortium through NASA National Space Grant NNX15AI19H. We thank the anonymous reviewer for constructive suggestions to improve the paper.